\documentstyle[twoside,fleqn,espcrc2]{article}
\input{epsf.sty}

\def\f{\phi}                    %       \varphi
\def\g{\gamma}

\def\j{\psi}
\def\k{\kappa}

\def\m{\mu}

\def\o{\omega}
\def\p{\pi}                     % Also, \varpi
\def\s{\sigma}                  %       \varsigma

\def\x{\xi}

\def\ce{{\cal E}}
\def\ch{{\cal H}}
\def\cz{{\cal Z}}

\def\svev#1{\left\langle #1\right\rangle}       % variable < >
\def\gtap{\raisebox{-.4ex}{\rlap{$\sim$}} \raisebox{.4ex}{$>$}}   % > or ~
\def\ltap{\raisebox{-.4ex}{\rlap{$\sim$}} \raisebox{.4ex}{$<$}}   % < or ~

\def\tk{\tilde\kappa}
\def\vh{v_{\rm H}}
\def\gdofs{\mbox{\it gdofs}}

\def\pslash{\rlap{\hbox{$\mskip 1 mu /$}}p}      % good slash for lower case

\def\NPB#1{Nucl. Phys. {\bf B#1}}
\def\NPBP#1{Nucl. Phys. (Proc. Suppl.) {\bf B#1}}
\def\PLB#1{Phys. Lett. {\bf B#1}}

\def\PRL#1{Phys. Rev. Lett. {\bf #1}}

\newcommand{\AmS}{{\protect\the\textfont2
  A\kern-.1667em\lower.5ex\hbox{M}\kern-.125emS}}

\title{Gauge-Fixing Approach to Lattice Chiral Gauge Theories, Part II}

\author{Wolfgang Bock,\address{Institute of Physics, 
        Humboldt University, \\ 
        Invalidenstr. 110, 10099 Berlin, Germany}%
        \thanks{presenters at LATTICE'97, Edinburgh}
        Maarten Golterman\address{Department of Physics,
        Washington University, \\
        St. Louis, MO 63130, USA} %
        and 
        Yigal Shamir\address{School of Physics and Astronomy, 
        Beverly and Raymond Sackler Faculty of Exact Sciences, \\
        Tel-Aviv University, Ramat Aviv 69978, Israel}$^*$}
       
\begin{document}

\begin{abstract}
In this more technical part we give additional details
on the gauge-fixing approach~\cite{rom,gf} presented in~\cite{bgs}.
We also explain how the gauge-fixing approach evades 
the Nielsen-Ninomiya~\cite{NNKS} no-go theorem.
\end{abstract}

\maketitle

\section{Introduction and Conclusion}

The gauge-fixing approach to lattice chiral gauge theories~\cite{rom,gf}
is presented in our first contribution to
these proceedings~\cite{bgs},  henceforth denoted by $I$. 
Here we discuss several issues in more detail.

Consider a continuum chiral gauge theory with a gauge group 
$G = {\rm U(1)}_L$ and a renormalizable gauge-fixing action, cf.\
eq.~(I.4). (We use (I.n) to denoted eq.~(n) in $I$.)
Now take the limit $g \searrow 0$, keeping the product 
$\tk \equiv (2\x g^2)^{-1}$ fixed. 
This {\it reduction} leads to a free higher-derivative action for the 
the gauge degrees of freedom (\gdofs) which are still propagating because 
$\tk$ is finite. (There are no Faddeev-Popov ghosts in the U(1) case.)
In addition, one obtains a set of free fermions 
in complex representations of U(1)$_L$, which are decoupled
from the \gdofs. 
This reduction is interesting for two purposes.
First, it provides us with a well-defined procedure for assigning 
the fermions to representations of the gauge group.
We can then inspect whether the fermion spectrum is vector-like or chiral.
It also tells us what the continuum limit (CL) should be,
if a similar reduction is applied to the lattice theory.

On the lattice, setting $g=0$ with fixed $\tk$ means that the link variables
are constrained to $U_{\m x} = \f_x^\dagger \f_{x+\m}$
in the action, and the integration measure 
$\prod dU_{\m x}$ is replaced by $\prod d\f_x$.
This defines the {\it reduced} model, which on the lattice is an interacting
theory involving the compact scalar field $\f_x \in {\rm U(1)}$.
We note that also the lattice momentum takes values on a compact space
(unlike its continuum counterpart).  
The lattice kinematics has the all-important dynamical
consequence, that the \gdofs\ represented by the
$\f_x$ field {\it couple} to fermions
even if we are careful to choose an anomaly-free spectrum.
This coupling has led to the failure of many chiral fermion proposals 
(for a review see~\cite{rev}),
which can be explained~\cite{gen} by the applicability of the Nielsen-Ninomiya
theorem~\cite{NNKS} to a wide class of models.

Here we consider the reduced model of a U(1)
lattice gauge theory with a Lorentz gauge-fixing action 
and a mass counterterm. For the \mbox{$g \ne 0$} gauge theory, 
the bosonic and fermionic lattice actions are respectively
eqs.~(I.8) and~(I.1). The reduced model's action
is eq.~(I.11), where $(2\x g^2)^{-1}=\tk$. 

Our main result is that the expected fermion spectrum 
is obtained in the CL of the reduced model.
It consists of charged left-handed (LH) and neutral right-handed (RH) fermions, 
which are free and decoupled from the {\it gdofs}.
This reproduces correctly the $g \searrow 0$ limit 
of the target gauge-fixed continuum theory. 
If we now take $0 < g \ll 1$, only the charged LH fermions will couple
to the gauge field, and so we  
may expect that a U(1) chiral gauge theory will emerge in the 
scaling region of the full lattice gauge theory,
provided the fermion spectrum is anomaly-free.

%%%% FIG. 1 %%%%%
\begin{figure}[t]
\vspace*{-.6cm}
\begin{tabular}{c}
\hspace*{-1.1cm} 
\epsfxsize=9.30cm
\epsfbox{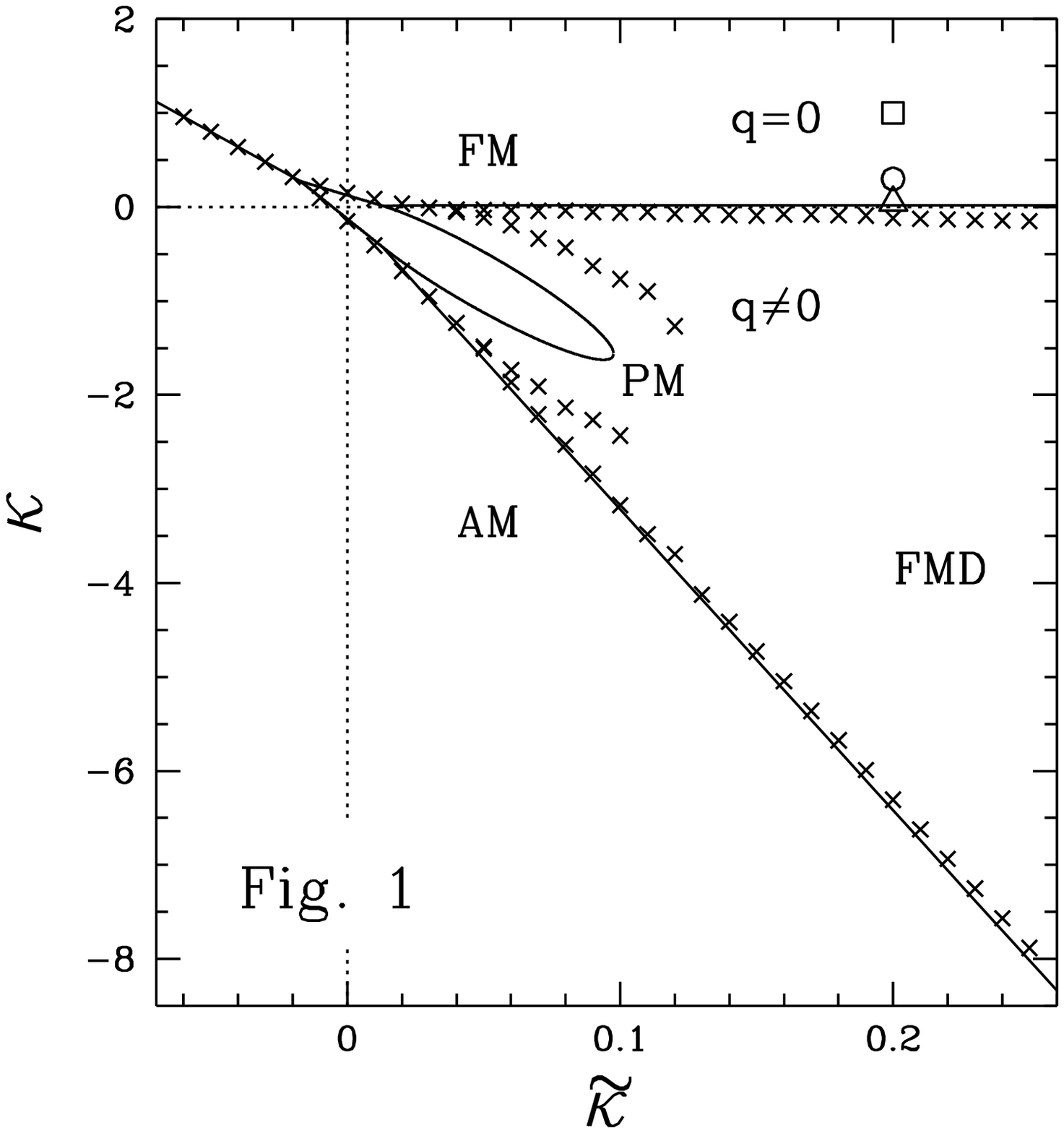}
\vspace*{-2.0cm}
\end{tabular}
%\caption{phase diagram}
%\label{autonum}
\end{figure}
%%%%%%%%%%%%%%%%%

%%%%%%%%%%%%%%%%%%%%%%%
\section{Phase Diagram}

In this section we discuss the phase diagram of the purely bosonic
part of the reduced model, cf.\ (I.11).
The novel feature of the phase diagram is the existence of an FMD
({\it ferromagnetic directional}) phase, characterized by the 
condensation of a space-time vector field. As discussed in $I$,
in the $g \ne 0$ theory one has $\svev{A_\m} \ne 0$
provided $\k$, the parameter of the mass counterterm, 
is smaller than $\k_c$. 
In the reduced model, the FMD phase is characterized by a vector
order parameter $0 < q_\m < 2\p$, $\m=1,\ldots,4$,  
and by a {\it helicoidal magnetization} $\vh$ defined by
\begin{equation}
  \svev{\f_x} = \vh \exp\Big( i\sum_\m q_\m x_\m \Big) \,.
\label{vh}
\end{equation}
Below we will ignore the global U(1) phase, and assume that $\vh=|\vh|\ge 0$. 
In the special case $q=(0,0,0,0)$ ($q=(\p,\p,\p,\p)$), $\vh$
coincides with the ordinary (staggered) magnetization, and we 
have an FM (AM) phase. For $\vh=0$ (and undefined $q_\m$) we have a PM phase.
The U(1) symmetry of the reduced, purely bosonic model is broken 
in the FM, AM and FMD phases. (If we include the fermions, 
the symmetry breaking is 
${\rm U(1)}_L \times {\rm U(1)}_R \to {\rm U(1)}_V$, 
where U(1)$_V$ is the diagonal subgroup.)
In the FMD phase, discrete
rotation invariance is also broken spontaneously.
The full global symmetry is restored on the FM-FMD line (Sect.~3).

In the mean-field (MF) approximation, $\f_x \sim \svev{\f_x}$ is
expressed in terms of $\vh$ and $q_\m$ using eq.~(\ref{vh}).
Following the standard steps of the MF calculation we obtain
the phase diagram depicted in Fig.~1 \cite{bos}. 
The MF phase transition (solid) lines
are all continuous, except the FM-AM transition which is first order. 
The crosses are the results of a numerical simulation on an $8^4$ lattice.
We see that the agreement between the MF and numerical results is very 
good, except  for the PM-FMD transition. Simulations
on different lattices suggest that the results for the PM-FMD
line could come closer to the MF prediction in the infinite volume limit.

%%%%%%%%%%%%%%%%%%%%%%%%%
\section{Continuum Limit}

In the $g \ne 0$ theory, the CL corresponds to 
$\svev{A_\m} = 0$ and a vanishing vector boson mass,
which is achieved by approaching the FM-FMD phase boundary.
(For U(1) we keep $0 < g \ll 1$ because of triviality.)
In the reduced model $q=0$ and $\vh=v > 0$ in the FM phase. 
We define the CL by approaching the FM-FMD line from the FM phase,  
i.e.\ $\k \searrow \k_c(\tk)$ at fixed $\tk$.
Based on the good agreement between perturbation theory (PT) and the 
numerical results discussed below, we believe
that for $\k \sim \k_c(\tk)$ the fermion spectrum 
at finite $\tk \;\gtap\; 0.1$ is the same as for $\tk \to \infty$.

Still within the purely bosonic theory, we now discuss the
critical behavior of the magnetization $v$ in the CL. 
As explained in $I$, PT is an expansion in $\tk^{-1} = 2\x g^2$.
At the classical level $v=|\svev{\f}|=1$. However,
this is a very poor approximation because of infrared effects.
The one-loop approximation is obtained by a gaussian integration,
which gives rise to eq.~(I.15). The infrared divergence of
the momentum integral in eq.~(I.15)
is cut off by $m^2=\k/\tk$.
The final one-loop result is $v \sim m^{2\eta}$ where 
$\eta=(64\p^2\tk)^{-1}$. The existence of a coupling-constant dependent critical
exponent resembles the spin-wave phase of the XY-model
in two dimensions.

%%%% FIG. 2 %%%%%
\begin{figure}[t]
\vspace*{-.6cm}
\begin{tabular}{c}
\hspace*{-1.1cm} 
\epsfxsize=9.30cm
\epsfbox{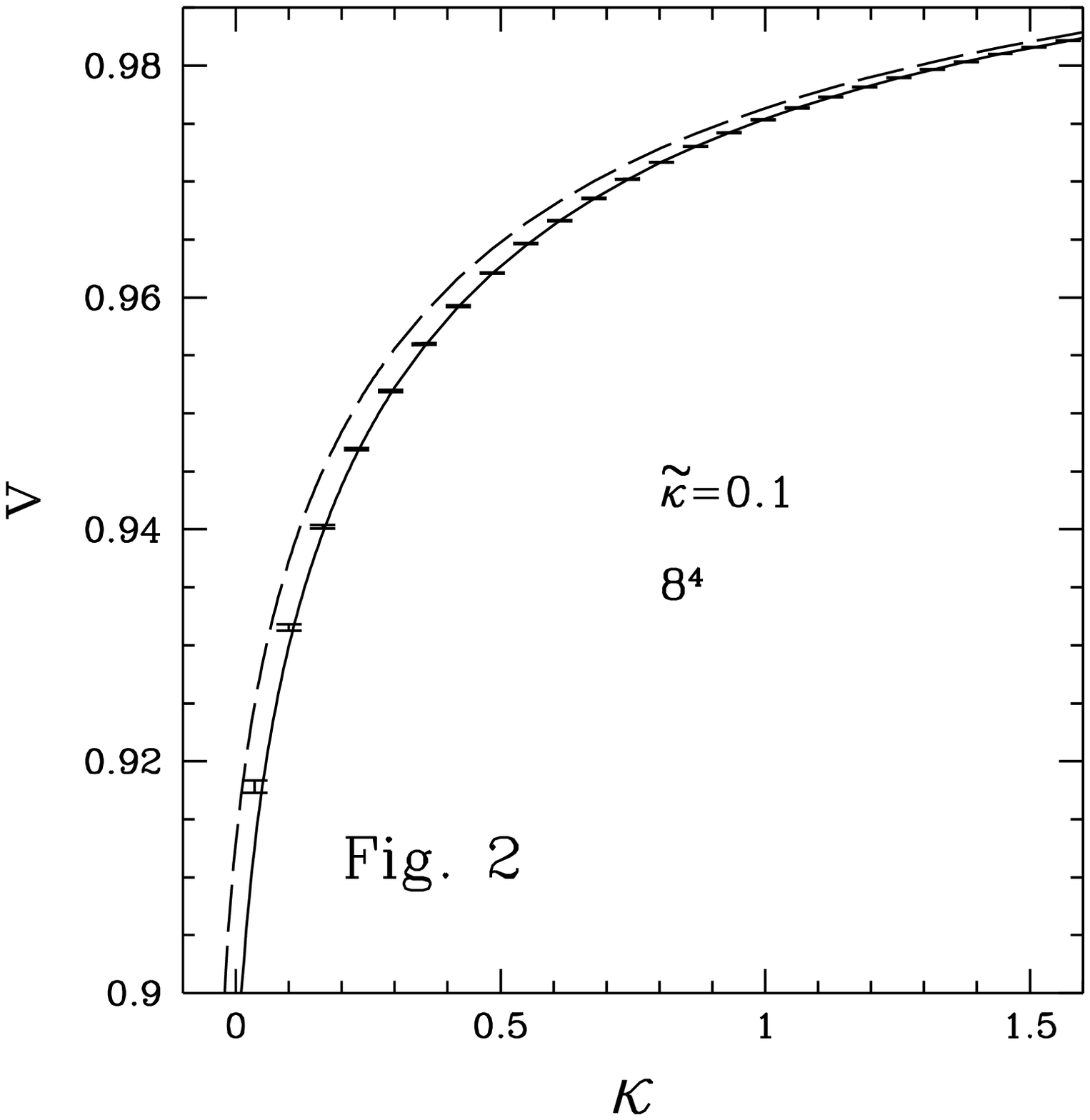}
\vspace*{-2.0cm}
\end{tabular}
%\caption{v(q)}
\end{figure}
%%%%%%%%%%%%%%%%%

In order to obtain a better prediction for the magnetization, we
first resum the one-loop corrections to the Goldstone field propagator
$G^{-1}_{\rm 1-loop}(p) = G^{-1}(p) + \Sigma(p)$. The self-energy 
$\Sigma(p)$ comes from a tadpole diagram (see~\cite{bos} for
the explicit expression). We obtain a two-loop
prediction for $v$ by replacing the tree-level propagator $G(p)$
in eq.~(I.15) with $G_{\rm 1-loop}(p)$. 
The self-energy calculation also leads to the one-loop estimate 
$\k_c = 0.02993+O(1/\tk)$ in the limit $\tk \to \infty$.

High precision data for the magnetization on an $8^4$ lattice at $\tk=0.1$
are shown in Fig.~2. The dashed and solid lines are respectively
the one-loop and two-loop results for $v$,
which were obtained by evaluating the lattice momentum sums numerically 
on the same lattice size. The finite lattice propagator does not contain
the zero momentum mode (associated with the global U(1) phase) 
that decouples from the action. We see that the two-loop
result is in very good agreement with the numerical simulation,
even at such a relatively small value of $\tk$. (This is because  
the actual expansion parameter is $1/(16\p^2\tk)$, not $1/\tk$.)
Results at a number of different volumes~\cite{bos} also show that 
the critical coupling approaches its infinite volume limit
from below as the lattice size in increased.

Finally we mention that the helicoidal magnetization $\vh$ also vanishes
with the same critical exponent $\eta$, when the FM-FMD line is approached
from the FMD phase. There is qualitative agreement between 
the analytic and numerical results in the FMD phase~\cite{bos}.
However, because of subtle finite-size effects ($q$ is quantized
on a finite lattice) the detailed behavior of the order parameters 
in the FMD phase, as well as very close to $\k_c$ in the FM phase, 
is not yet fully understood.

%%%%%%%%%%%%%%%%%%%%%%%%%%
\section{Fermion Spectrum}

As discussed above, the full global symmetry 
is restored in the limit $\k \to \k_c$.
Our aim now is to determine the U(1)$_L\times$U(1)$_R$
quantum numbers of the massless fermions for $\k=\k_c$.
Eq.~(I.17) is the reduced model's fermion action in the {\it neutral}
formulation, which involves the
four-component neutral field $\j^n$ transforming only under U(1)$_R$.
The substitution $\j^n \to \f^\dagger\j^c$ leads to the {\it charged}
formulation, where the charged field $\j^c$ transforms only 
under U(1)$_L$. In both formulations PT is manifestly infrared finite,
because the Goldstone field $\theta$ has only derivative couplings.

For definiteness we consider here the charged fermion propagator
\begin{equation}
  S^c_{\rm 1-loop}(p) = [S^{-1}(p) + \Sigma^c(p)]^{-1} \,.
\label{sc} 
\end{equation}
The tree-level propagator $S(p)$ is the free massless
Wilson propagator. Explicitly,
$S^{-1}(p)= \sum_\m \{ i\g_\m \sin p_\m +2r \sin^2 {p_\m\over 2} \}.$
The explicit expression for the one-loop self-energy 
in the charged formulation $\Sigma^c(p)$ will be given elsewhere~\cite{pth}.

First, $S^c_{\rm 1-loop}(p)$
is regular and $O(1)$ if one or more momentum components are close to $\p$.
This implies that the doublers (situated at the fifteen corners
of the four-dimensional Brillouin zone) have $O(1/a)$ masses, and therefore
decouple. 

%%%% FIG. 3a %%%%%
\begin{figure}[t]
\vspace*{-.6cm}
\begin{tabular}{c}
\hspace*{-1.1cm} 
\epsfxsize=9.30cm
\epsfbox{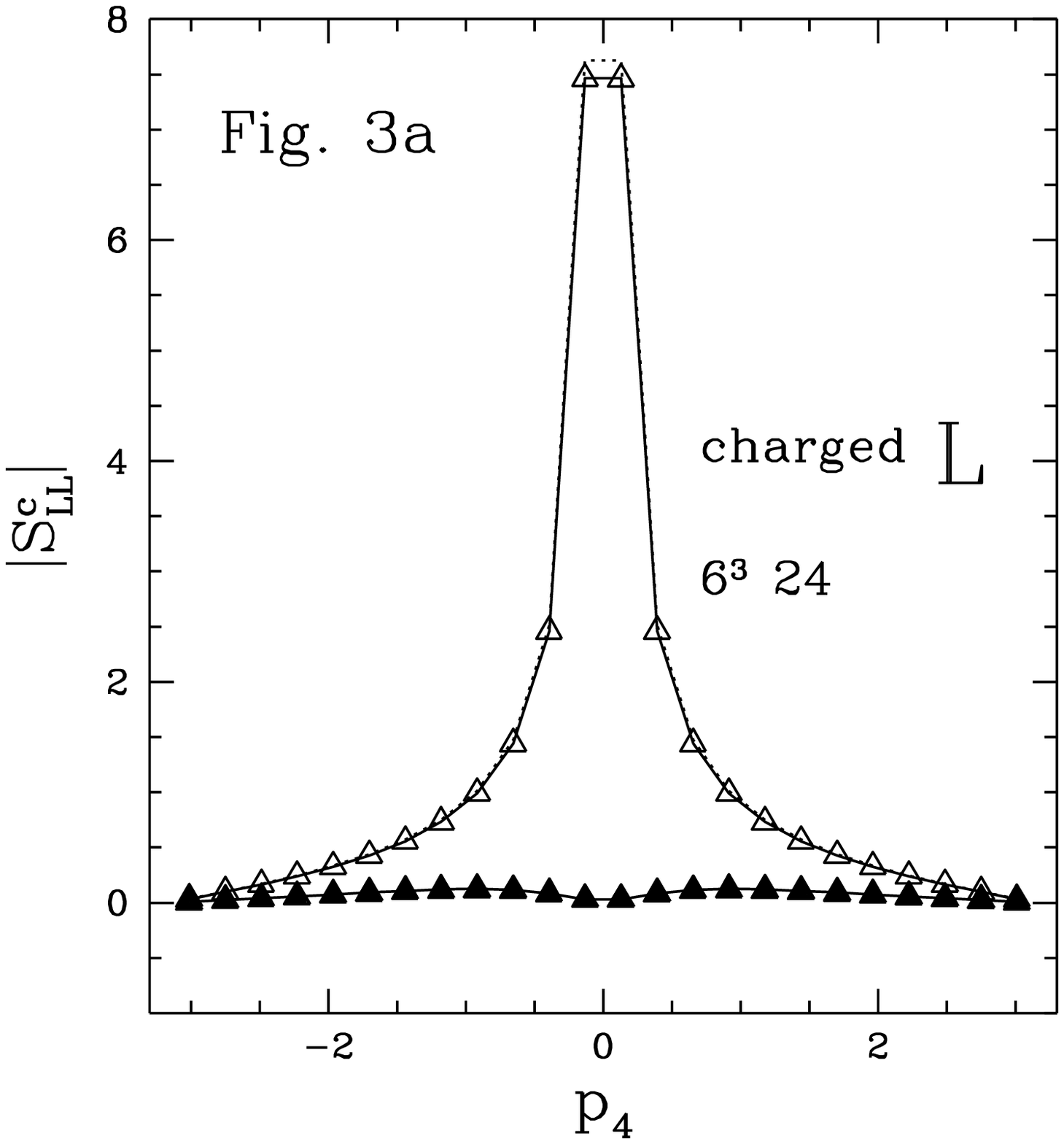}
\vspace*{-2.0cm}
\end{tabular}
%\caption{cLd}
\end{figure}
%%%%%%%%%%%%%%%%%

%%%% FIG. 3b %%%%%
\begin{figure}[t]
\vspace*{-.6cm}
\begin{tabular}{c}
\hspace*{-1.1cm} 
\epsfxsize=9.30cm
\epsfbox{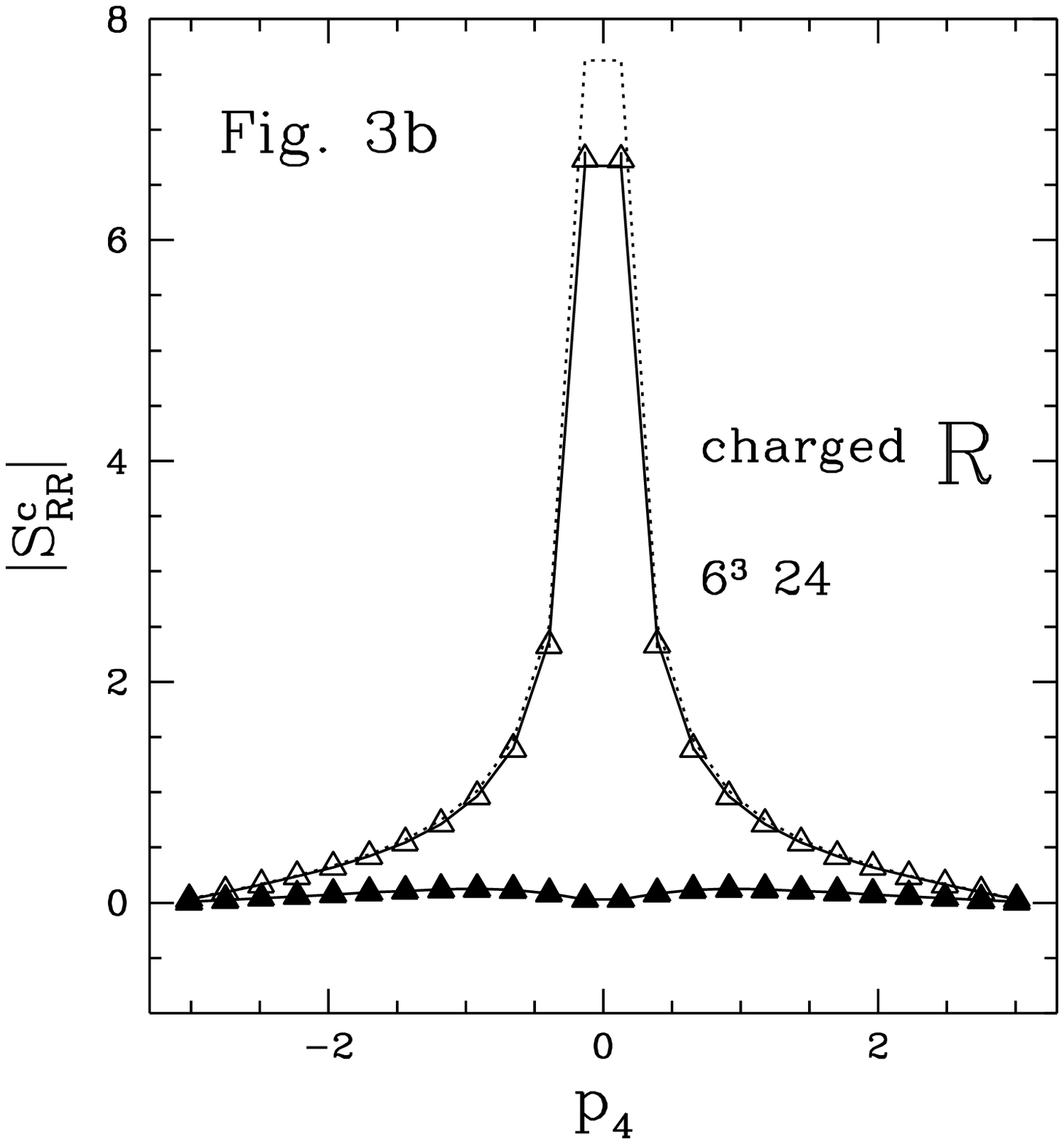}
\vspace*{-2.0cm}
\end{tabular}
%\caption{cRd}
\end{figure}
%%%%%%%%%%%%%%%%%

We next consider the small momentum behavior of the propagator.
Substituting $p=0$ we obtain  $\Sigma^c(0) = 0$, which is a consequence
of a shift symmetry~\cite{gps}, cf.\ eq.~(I.18).
This implies that no fermion-mass counterterm is needed.
Finally we calculate the small-$p$ nonanalytic behavior of $\Sigma^c(p)$.
In the limit $\k \searrow \k_c(\tk) \sim 0$ we find
\begin{equation}
 \Sigma^c(p) \approx -i (32\p^2\tk)^{-1}\, \pslash\, P_R\, \log(p^2) \,.
\label{sigma}
\end{equation}
We see that nonanalytic terms occur only in the RH charged channel!
While the massless tree-level LH pole remains isolated, 
in the RH charged channel we now have a cut.
Going back to configuration space, we find for the long-distance part
\begin{equation}
  S^c_{RR}(x,y) \approx S^n_{RR}(x,y) \svev{\phi(x) \phi^\dagger(y)} \,,
\label{fctr}
\end{equation}
where $S^{c,n}_{RR}(x,y) = \svev{\j_R^{c,n}(x) \bar\j_R^{c,n}(y)}$.
Since $\j^c_R=\phi\j^n_R$, eq.~(\ref{fctr}) means
that the RH charged propagator {\it factorizes}.
A complementary situation is found for the neutral propagator, 
which has an isolated pole in the RH channel and a cut
in the LH channel. The LH neutral propagator $S^n_{LL}(x,y)$
factorizes as $S^c_{LL}(x,y)\svev{\phi^\dagger(x) \phi(y)}$.
Together, this gives strong evidence that the spectrum consists of
charged LH and neutral RH fermions, 
which are free and decoupled from the unphysical \mbox{\it gdofs} in the CL. 
(This generalizes to an arbitrary choice of the fermion species,
and is consistent with the vanishing of the anomaly in
the absence of a transversal gauge field.)

In $I$ we have compared one-loop PT with simulation results for 
the neutral propagator in momentum space. Here we present
the complementary results for the charged propagator.
All data were obtained in the quenched approximation 
on a $6^3\times 24$ lattice at $\tk=0.2$ and $r=1$, cf.\ eq.~(I.1).
For the fermion fields we use p.b.c.\ in $x_1,x_2,x_3$ and a.p.b.c.\ in $x_4$.
For the scalar field we use p.b.c.\ in all directions.
The scalar field configurations were generated 
with a 5-hit Metropolis algorithm, and 4000 configurations were skipped 
between two successive fermionic measurements. We inverted the fermion matrix 
on a total of 50 scalar field configurations. 
In Fig.~3 we plot the modulus of $S^c_{LL}$ and $S^c_{RR}$ 
as a function of $p_4$, 
for $\vec{p}=(0,0,0)$ (open triangles) and $\vec{p}=(\p,0,0)$
(filled triangles) at $\k=0.05$, where the charged
propagator is now
\begin{equation}
  S^c = -i [ S^c_{LL}(p_4) P_L + S^c_{RR}(p_4) P_R ] \g_4 + S^c_{LR}(p_4). 
\label{scp}
\end{equation}
For comparison we also show the tree-level
(dotted line) and one-loop (solid line) results, which were obtained
for the same lattice size. We see that $S^c_{LL},S^c_{RR} \sim 0$
if one or more momentum components are close to $\p$.
This confirms the decoupling of the doublers discussed earlier.

Both $S^c_{LL}$ and $S^c_{RR}$ diverge for $p \to 0$.
In order to determine which propagator has a pole,
the data at $\vec{p}=(0,0,0)$ are replotted in Fig.~4, which shows
the ratios $S^c_{LL}/S_{LL}$ and $S^c_{RR}/S_{RR}$ as a function of $p_4$.
Results are shown for three $\k$-values, 1 (squares), 0.3 (circles) and 0.05
(triangles), which are decreasing towards $\k_c$.
(These points are marked in the phase diagram Fig.~1.)
Fig.~4a for $S^c_{LL}$ is qualitatively the same
as Fig.~2a in $I$ for $S^n_{RR}$.
In both cases we find that the tangent is parallel to the abscissa
at $p_4=0$ for all values of $\k$.
Hence $S^c_{LL}$ and $S^n_{RR}$ have a pole at $p=0$.
The wave-function renormalization constant can be read off from the intercept 
at $p_4=0$. We find $Z_R=1$ for the RH neutral field
(which is a consequence of the shift symmetry~\cite{gps}) and $Z_L \ltap 1$ 
for the LH charged field. A completely different behavior is found in
Fig.~4b for $S^c_{RR}$ which, similarly to $S^n_{LL}$ (see $I$),
shows a pronounced dip at $p_4 \sim 0$. 
We expect this dip to approach the logarithmic singularity 
in eq.~(\ref{sigma}) in the infinite volume limit for $\k \searrow \k_c$.
The solid lines represent again one-loop PT. 
The data for the propagator ratios is in good agreement with PT,
and it is very likely that the small discrepancies are due to higher-order
effects.

%%%% FIG. 4a %%%%%
\begin{figure}[t]
\vspace*{-.6cm}
\begin{tabular}{c}
\hspace*{-1.1cm} 
\epsfxsize=9.30cm
\epsfbox{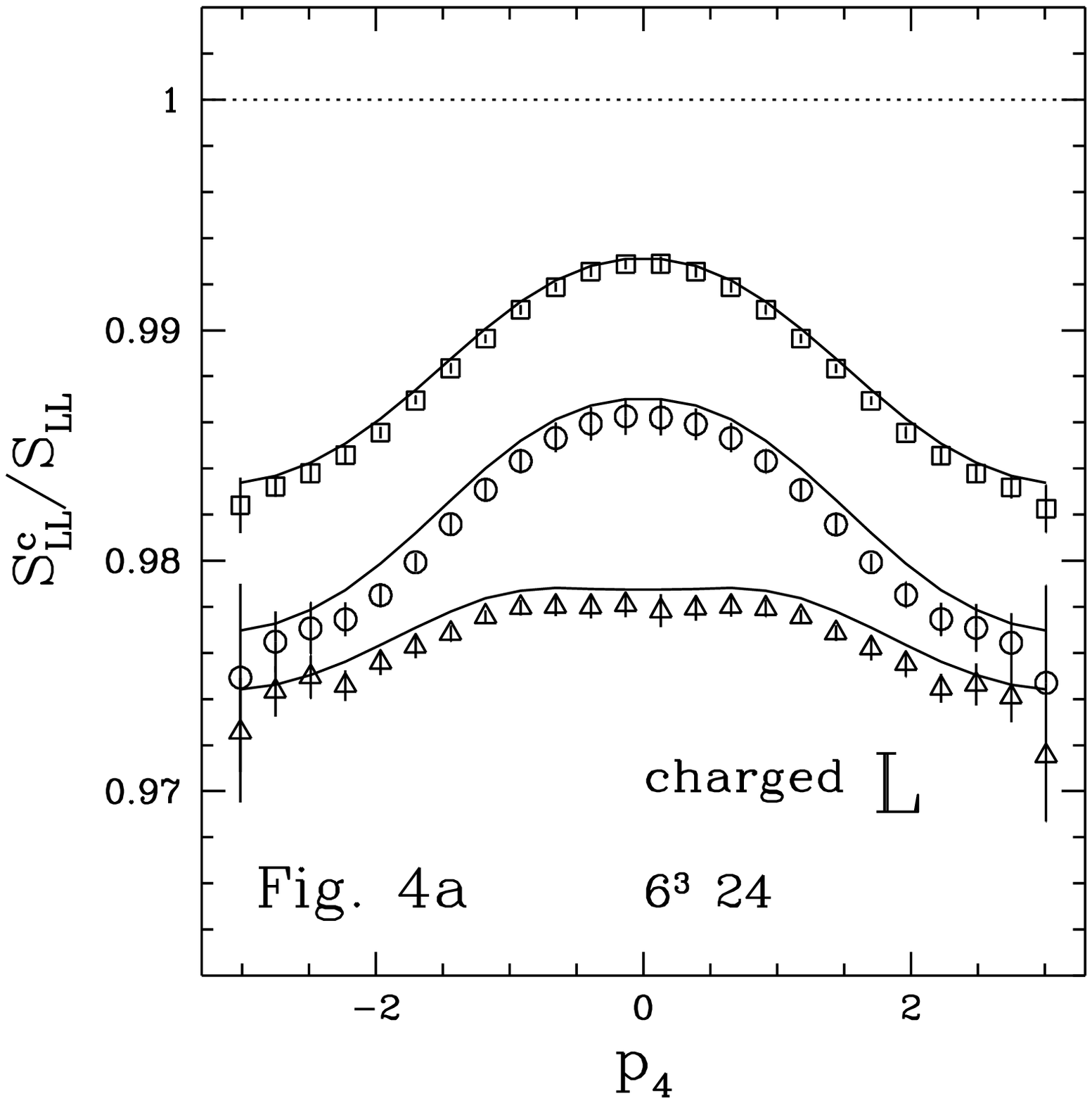}
\vspace*{-2.1cm}
\end{tabular}
%\caption{cL}
\end{figure}
%%%%%%%%%%%%%%%%%

%%%%%%%%%%%%%%%%%%%%%%%%%%%%%%%%%%%%
\section{Evading the No-Go Theorems}

In a nut-shell, the Nielsen-Ninomiya (NN) theorem~\cite{NNKS} asserts that
if the lattice action is local, there is an equal number of LH and RH 
fermions in any given representation.
The NN theorem is formulated for free lattice theories, but
it is also applicable to a wide range of interacting 
theories~\cite{gen}. For example (see $I$) any lattice
gauge theory with a gauge-noninvariant action can be reformulated
as a gauge invariant lattice theory with an additional group-valued
scalar field $\f_x \in G$, which is associated with the \gdofs.
Setting $U_{\m x}=I$ in the new action gives rise to the reduced
model, where the local symmetry $G$ turns into a
global one. The NN theorem is now relevant provided 
$G$ is not broken spontaneously.

%%%% FIG. 4b %%%%%
\begin{figure}[t]
\vspace*{-.6cm}
\begin{tabular}{c}
\hspace*{-1.1cm} 
\epsfxsize=9.30cm
\epsfbox{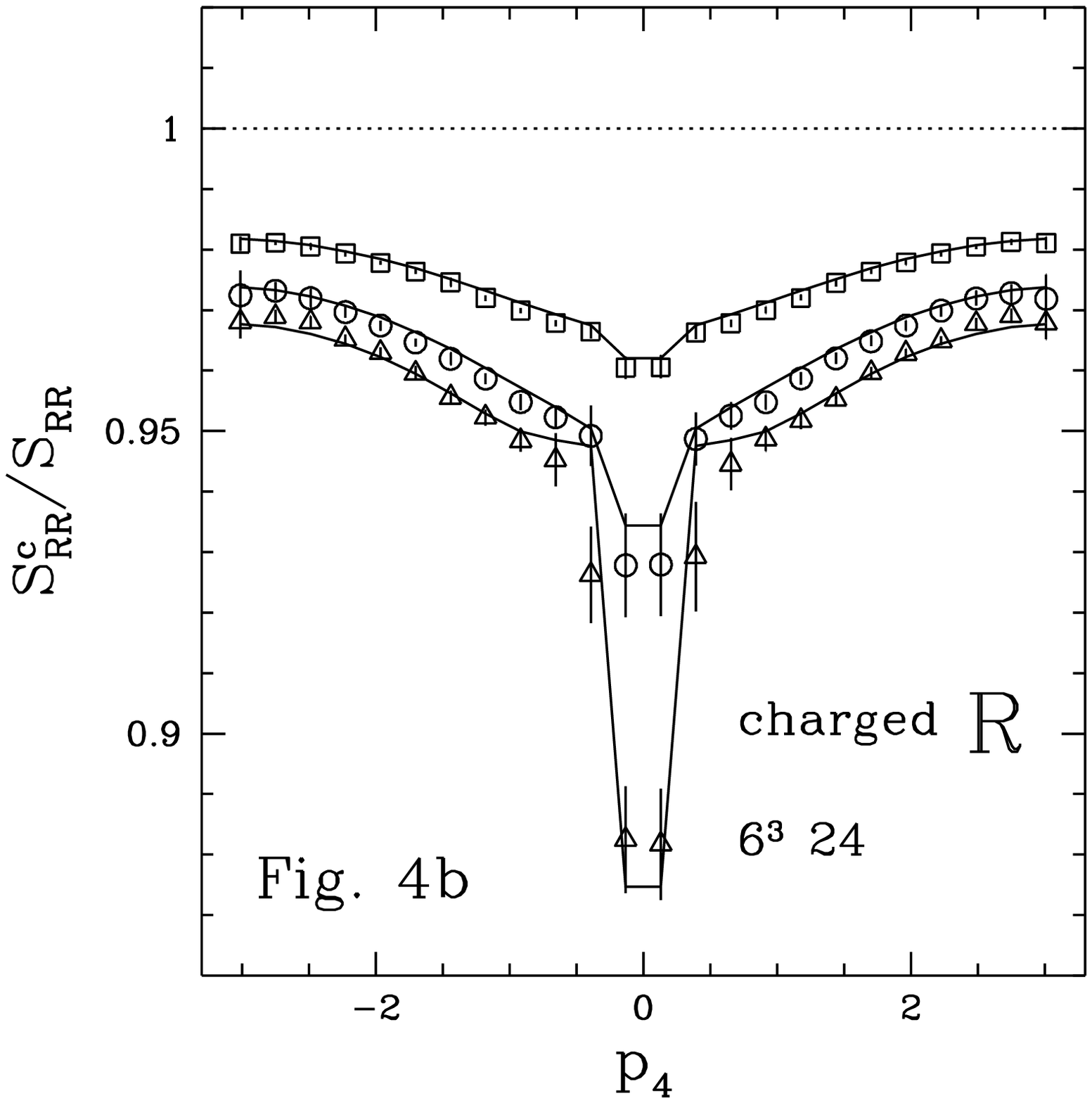}
\vspace*{-2.1cm}
\end{tabular}
%\caption{cR}
\end{figure}
%%%%%%%%%%%%%%%%%

  In our model $G={\rm U(1)}_L$ is not broken in the PM phase, as well as
on the FM-FMD line. While it does not seem possible to define
a chiral gauge theory in the PM phase~\cite{gen,rev}
or close to its boundary,
we do find (free) chiral fermions on the FM-FMD line.
We will now explain how the fermion two-point functions of the reduced model
can be understood in the framework of the NN theorem.

  Considering for definiteness the charged propagator $S^c(p)$, we define
\begin{equation}
  \ch^c(\vec{p}) = \g_4 [S^c(\vec{p},p_4=0)]^{-1} \,.
\label{h}
\end{equation}
The matrix $\ch^c(\vec{p})$ is hermitian. 
(However, $\ch^c(\vec{p})$ is not a hamiltonian, see below.)
Let us now rotate the lattice used in Sect.~4 such that its long side 
lies along the $x_3$-direction. The results for $S^c(0,0,0,p_4)$
turn into results for $S^c(0,0,p_3,0)$, which determine 
$\ch^c(p_3)=\ch^c(0,0,p_3)$ using eq.~(\ref{h}). 
$\ch^c(p_3)$ commutes with $\s_{12}=-i\g_1\g_2$. 
In the chiral representation
\begin{equation}
  \g_4 = \left(\begin{array}{cc}
         0 & I \\
         I & 0 
         \end{array}\right),\quad    
  \g_k = \left(\begin{array}{cc}
         0 & -i\s_k \\
         i\s_k & 0 
         \end{array}\right),
\end{equation}
$\g_5=\g_1 \g_2 \g_3 \g_4$, and the two-by-two spin-up ``subhamiltonian'' reads 
\begin{equation}
  \ch^c_\uparrow = 
  \left(\begin{array}{cc}
  S^c_{RR} & S^c_{LR} \\
  S^c_{LR} & -S^c_{LL}
  \end{array}\right)^{-1} \,.
\label{22}
\end{equation}
The eigenvalues $\ce^c_\pm$ of $\ch^c_\uparrow$
are depicted in Fig.~5 as a function of $p_3$, using our data for $\k=0.05$. 
As before, the solid (dotted) lines are
the one-loop (tree-level) results.

%%%% FIG. 5 %%%%%

\begin{figure}[t]
%\vspace*{-.6cm}
\begin{tabular}{c}
\hspace*{-1.1cm} 
\epsfxsize=9.30cm
\epsfbox{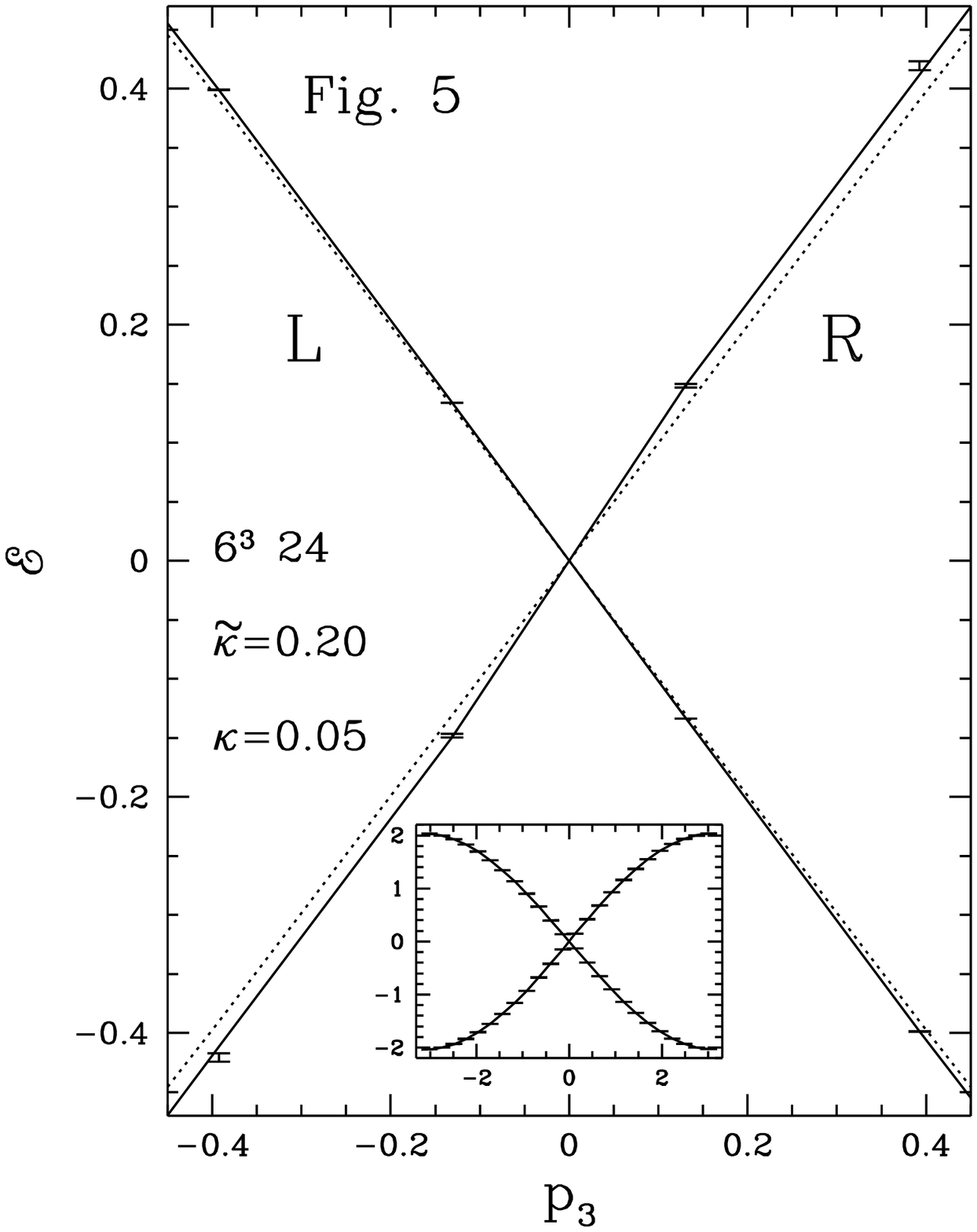}
\vspace*{-1.0cm}
\end{tabular}
%\caption{cR}
\end{figure}
%%%%%%%%%%%%%%%%%

We first consider the free Wilson case (dotted lines).
Here $\ch(\vec{p}) = \g_4 [S(\vec{p},0)]^{-1}$ coincides with the lattice
hamiltonian. For $\vec{p}=(0,0,p_3)$, the two spin-up eigenvalues 
are $\ce_\pm = \pm p_3$ in the limit $p\to 0$. 
These describe a positive-helicity (RH)
and a negative-helicity (LH) fermion, which together
constitute a Dirac fermion.

We now turn to our model. The small figure shows that 
the two eigenvalues form a continuous curve 
that doubly covers the periodic Brillouin zone. 
As in the free Wilson case, this is a consequence of locality.
$\ch^c_\uparrow$ has a positive-slope and a negative-slope zero. 
Technically, this is in agreement with the NN theorem. Nevertheless,
our model evades the negative physical conclusion of the theorem.
For $p \to 0$, $\ce^c_- = - Z_L^{-1} p_3$ is linear.
We can absorb the $\j^c_L$ renormalization constant via
$\ch^c \to \ch' = \cz \ch^c \cz$ with
$\cz=Z_L^{1\over 2} P_L + P_R$. 
For $\ch'_\uparrow$ we obtain the relativistic relation
$\ce'_- = - p_3$ in the limit $p \to 0$, which corresponds to a
(charged) negative-helicity fermion.

The positive-slope eigenvalue $\ce^c_+$ in Fig.~5 is {\it not} linear:
$|\partial \ce^c_+ / \partial p_3|$ increases for small $|p|$.
$\ce^c_+$~is consistent with $S^c_{RR} \sim P_R\, \pslash^{-1}\, |p|^{4\eta}$ 
(and a divergent slope $\partial \ce^c_+ / \partial p_3$ for $p \to 0$) 
in the infinite volume limit for $\k \searrow \k_c$.
This result is derived from the factorization formula eq.~(\ref{fctr})
where $\svev{\f(x)\f^\dagger(y)} \sim |x-y|^{-4\eta}$.
As discussed before, it is consistent with 
the existence of a {\it neutral} positive-helicity fermion. 
(The deviations from linearity of the data points
are statistically significant. Their smallness 
is explained by the smallness of the critical exponent $\eta$.)

If, instead of setting $p_4=0$, we take the CL
and analytically continue $p_4 \to i\o$,
we find in the LH charged channel a pole
$Z_L (\o+\vec\s\cdot\vec{p})^{-1}$.
In this limit $\ch'_{LL} = Z_L \ch^c_{LL}$ coincides with the LH
Weyl hamiltonian $-\vec\s\cdot\vec{p}$.
This is not the case in the RH channel because of the
$|p|^{4\eta}$ factor. Therefore $\ch^c_{RR}$
is not a single-particle hamiltonian. 
This is a consequence of the infrared
singularity $\sim 1/(p^2)^2$ of the Goldstone field propagator 
for $\k \searrow \k_c$, for which no particle interpretation is known.
Fortunately, the physical fermions decouple from the unphysical sector.  
Thus, for the first time, we obtain free chiral fermions 
in the CL of a four-dimensional lattice model.

%%%%%%%%%%%%%%%%%%%%%%%%%%%%%%%%%%%%%%%%%%

\end{document}